\documentclass{svproc}
\usepackage{url}

\usepackage{amsmath}
\usepackage{amssymb}

\usepackage{amsfonts}
\newcommand{\ket}[1]{|{#1}\rangle}			
\newcommand{\bra}[1]{\langle{#1}|}

\usepackage{graphicx} 
\usepackage{xcolor}

\begin{document}
\mainmatter              
\title{Multicast quantum network coding as optimal symmetric universal cloning over a quantum
network}
\titlerunning{Multicast quantum network coding}  
%
\author{Go Kato\inst{1} \and Mio Murao\inst{2}\inst{,3}
 \and Masaki Owari\inst{4}}
\authorrunning{Go Kato et al.} 
%
\tocauthor{Go Kato, Mio Murao, Masaki Owari}
\institute{Advanced ICT Research Institute, NICT 4–2–1, Nukui-Kitamachi, Koganei,\\
Tokyo 184-8795, Japan,
\and
Department of Physics, Graduate School of Science, The University of Tokyo, Hongo 7-3-1, Bunkyo-ku, Tokyo 113-0033, Japan,
\and
Trans-scale Quantum Science Institute, The University of Tokyo, Bunkyo-ku, Tokyo 113-0033, Japan
\and Department of Computer Science, Faculty of Informatics, Shizuoka
University, Chuo-ku, Hamamatsu-shi, Shizuoka 432-8011, Japan\\
\email{masakiowari@inf.shizuoka.ac.jp}}

\maketitle              

\begin{abstract}
We study the problem of perfectly multicasting symmetric universal clones of unknown quantum states over quantum networks with free classical communication.
 We construct a protocol that multicasts symmetric universal clones of input states from multiple source nodes by extending the quantum network coding protocol proposed by Kobayashi et al.  
We further establish a sufficient condition for perfect multicast in the single-source setting. Specifically, we show that when a single copy of a $q^r$-dimensional input state is available at the source node, where $q$ is a sufficiently large prime power, perfect multicast of the corresponding symmetric universal clone is achievable using a small amount of entanglement shared among the target nodes.
 This result holds for quantum networks represented by an undirected graph $G$, where each edge corresponds to a noiseless $q$-dimensional quantum channel, provided that there exists an acyclic directed graph $G'$ obtained by assigning directions to the edges of $G$ such that the minimum cut of $G'$ is at least $r$.

\end{abstract}
\section{Introduction}

The max-flow min-cut theorem of flow networks states that the maximum flow from a source
node to a terminal node is equal to the size of a minimum cut on
a network. Alswede et al. extended this max-flow min-cut theorem to
multicast network communication \cite{Ahlswede2000}: Suppose a single
sender (or a source node) $s$ tries to send identical information to multiple receivers (or terminal nodes) $t_{1},\cdots,t_{n}$ on a classical network under the condition that intermediate nodes can apply linear or non-linear operations on accepted packets before sending them to the next nodes. 
The maximum amount of information which can be multicasted from $s$ to all the target nodes is equal to the minimum with respect to $i$ of the minimum cut between $s$ and $t_{i}$ when the alphabet used in the network communication is sufficiently large. 
This significant result opened the door to a new field, so-called, network coding, where coding on intermediate nodes on a network is studied to increase information flow, security, and robustness \cite{Yeung2007,HL2008}.

A quantum network consists of spatially separated quantum computers that are interconnected by quantum communication channels and classical communication channels \cite{Kimble2008,Rodney2014,Pirandola2017,Rigovacca_2018,Flamini2019,Hahn2019,Bauml2020,PhysRevA.106.L010401,doi:10.1126/science.aam9288}. Such networks enable the transmission, processing, and distribution of quantum information over long distances. In quantum networks, information is carried by quantum states, and a fundamental resource that distinguishes them from classical networks is quantum entanglement \cite{RevModPhys.81.865}. Entanglement describes quantum correlations between spatially separated systems that cannot be generated by local operations and classical communication (LOCC) alone. In particular, entanglement enables entanglement-assisted communication, where pre-shared entangled states enhance the ability to transmit quantum information. 
A basic example is quantum teleportation \cite{Bennett1993}, which shows that one qubit of quantum information can be faithfully transmitted from a sender to a receiver using shared entanglement and classical communication.
In particular, teleportation requires one shared maximally entangled bipartite state, where each party holds a two-dimensional quantum system (one ebit), together with two bits of classical communication.
From an operational viewpoint, teleportation implies that shared entanglement, combined with classical communication and local quantum operations, can effectively simulate a noiseless quantum channel. Consequently, the distribution and manipulation of entanglement play a central role in determining the communication capabilities of quantum networks.

Recently, in response to the rapid development of quantum network technology, 
a quantum analogue of network coding, called quantum network coding (QNC), has been studied in quantum information theory.
In analogy with classical network coding, QNC investigates whether appropriate processing of quantum states and entanglement at intermediate nodes can overcome network bottlenecks. In particular, by exploiting entanglement-assisted communication and LOCC-based operations at intermediate nodes, it is possible in certain network topologies to increase throughput and improve security beyond what is achievable by routing-based quantum communication alone \cite{Hayashi2007,Kobayashi2009,Kobayashi2010,Kobayashi2011,Leung2010,KOM2018,KOM2018a,AM2016,SKTM2011,SH2018,PhysRevA.76.040301,LXCQNY2017,MSNV2017,EKB2016,SINV2016,SLI2012,BR2014,Xu2015,10485002,PhysRevA.109.062606,Li2018,Zhang2022,Pan2021,LI20212651,Zhou2024,Shang2021,10404622,9427210,wang2023efficientsecurequantumnetwork,Liu2025,app12126163}.

Although multicast network coding is mainly studied in classical information theory, most existing research on QNC has focused on the multiple-unicast setting.
In multiple-unicast network coding, there are source nodes $s_{1},\cdots,s_{n}$ and terminal nodes $t_{1},\cdots,t_{n}$ on a network, and the goal is to transmit information from $s_i$ to $t_i$ for all $i$. 
One reason why the multicast setting is less explored in the quantum case is the no-cloning theorem, which prohibits perfect copying of an unknown quantum state \cite{WZ82}. 
As a result, perfect multicast of a single quantum state over a quantum network is impossible, and a direct quantum analogue of classical multicast network coding cannot be formulated. 
This situation contrasts sharply with the multiple-unicast case, where the quantum problem setting can be regarded as a direct extension of the classical one.

For the above reason, multicast QNC needs to have a purpose that is different from perfect multicast of a single copy of an unknown quantum state. One of the possible problem settings is considering \emph{perfect
distribution} of a quantum state over a quantum network \cite{Shi2006}.
To distribute a copy of a state $\ket{\Psi}\in\mathbb{C}^{d}$ from
a single source node to all $N$ terminal nodes, the source node must initially have $N$-copies of $\ket{\Psi}\in\mathbb{C}^{d}$. Shi et al. considered this problem and showed that $\ket{\Psi}^{\otimes N}\in\boldsymbol{\left(\mathbb{C}^{d}\right)^{\otimes N}}$ can be sent from one node to another by just sending the symmetric subspace of $\left(\mathbb{C}^{d}\right)^{\otimes N}$, whose dimension $d\left[N\right]:=\frac{\left(d+N-1\right)!}{N!\left(d-1\right)!}$
is much smaller than the dimension of $\left(\mathbb{C}^{d}\right)^{\otimes N}$.
Hence, we can increase the throughput of networks by this protocol. 

Kobayashi et al. treated the situation where a sender on a quantum
network initially has a single copy of a state $\ket{\Psi}:=\sum_{i=0}^{d-1}\alpha_{i}\ket{i}$
\cite{Kobayashi2010}. Their purpose is to produce a GHZ-type state
$\sum_{i=0}^{d-1}\alpha_{i}\ket{i}_{1}\otimes\cdots\otimes\ket{i}_{N}$
on terminal nodes, where the $i$-th system is on the $i$-th terminal
node. From this state, it is easy to reconstruct $\ket{\Psi}$ on
an arbitrary terminal node. In paper \cite{Kobayashi2010}, Kobayashi
et al. presented a way to construct a protocol achieving the above
task from a given classical multicast network code under the assumption that
we can freely use classical communication over a quantum network.
Here, the quantum network needs to be represented by the same graph
which represents the corresponding classical network, and each quantum
channel on the quantum network needs to have the same capacity of
the corresponding classical channel on the classical network.

Further, Xu et al. considered quantum network communication over a quantum network with multiple source nodes $s_1,\cdots, s_L$ and multiple terminal nodes $t_1,\cdots, t_N$, where initial states $\ket{\Psi_j}=\sum _{i=0}^{d-1}\alpha_{ij}\ket{i}$ are given in source node $s_l$ for all $l$. The goal of their network communication protocol is to construct the state $\ket{\phi}:=\sum_{i=0}^{d-1}f_k(\alpha_{1,k},\cdots, \alpha_{N,k})\ket{k}$ on each terminal nodes for  given functions $f_k: \mathbb{C}^N\rightarrow \mathbb{C}$ \cite{Xu2015}.  
Recently, Pan et al also proposed a protocol to probabilistically send a copy of a given quantum state $\ket{\psi}$ or an orthogonal complement of $\ket{\psi}$ to multiple terminal nodes on a quantum network
\cite{Pan_2022}.

Although the above protocols achieve multicast communication over quantum networks, it is slightly difficult to immediately accept them as quantum extensions of classical multicast network coding. 
First, the purposes of Xu et al's protocol and Pan et al.'s protocol are much more complicated in comparison to the simple multicast of quantum information to terminal nodes.  
Second, Shi et al's protocol does not implement any non-trivial coding, but it just throws away the orthogonal complement of a symmetric tensor product space of $\left(\mathbb{C}^{d}\right)^{\otimes N}$, which is generated
by all candidates of input states $\left\{ \ket{\Psi}^{\otimes N}\ |\ \ket{\Psi}\in\mathbb{C}^{d}\right\} $.
Further, we should note that Shi et al's protocol does not use classical
communication at all. On the other hand, as we can observe from research in a multiple-unicast setting, it is difficult to construct efficient
QNC without classical communication \cite{Hayashi2007}. As we have mentioned above, Kobayashi et al's protocol achieves multicast communication by outputting a GHZ-type state. However, as it is well known that this is not a good output state, when we consider this multicast protocol as a cloning machine.
For example, when an input state is $\ket{+}$ or $\ket{-}$, a reduced density matrix of the output is the completely mixed state, which does not have any information about the input state.

In this paper, we present a multicast protocol on a quantum network that may be accepted as a more straightforward extension of classical multicast network coding to a quantum network. Here, multicast
of an optimal universal clone of an input state \cite{BH1996,GM1997,BH1998,Werner1998} is considered as the best
multicast communication which is permitted by quantum mechanics. Hence,
we treat multicast-network coding on a quantum network as a protocol that perfectly sends optimal universal clones of an input state given at a source node
to terminal nodes. Here, for simplicity, we focus on symmetric universal cloning
and do not treat asymmetric universal cloning in this paper. Our interest especially
focuses on the relation between the size of the minimum cuts of quantum networks and the dimension of the input system whose optimal symmetric clones can be multicasted to terminal nodes. 
Kobayashi et al.'s quantum multicast network coding for sharing GHZ-type states achieves the minimum cut rate. Our results demonstrate that multicast of optimal symmetric universal clones  is possible at the minimum cut rate using Kobayashi et al.'s protocol and additional LOCC over a quantum network, provided a small amount of entanglement is shared at the terminal nodes.

Here, we should note the following fact about quantum and classical networks. When free classical communication is available on a quantum network, even if the direction of a quantum communication channel on the network is fixed beforehand, a quantum teleportation \cite{Bennett1993} achieves quantum communication in the opposite direction. Hence, in this case, a quantum network should be represented by an undirected graph $G$ \cite{Leung2010}. 
On the other hand, a situation will be changed when we consider a multicast classical network coding on a classical network represented by an undirected graph $G$, that is, each classical channel can be used either for forward communication or backward communication. 
In this case, since each channel is used for either forward or backward communication in classical network coding, the direction of the communication will be fixed. 
Hence, classical communications in the classical network coding is represented by a directed graph $G'$ derived by adding a direction to each edge of the undirected graph $G$. 

In this paper, we study multi-source multicast communication on a quantum network, where multiple source nodes independently perform multicast communication to a common set of terminal nodes. The single-source multicast scenario is included as a special case in which the number of source nodes is one.

We summarize our problem setting as follows:
\begin{enumerate}
\item There are $L$ source nodes ($s_{1},\cdots,s_{L}$) and $N$ terminal nodes ($t_{1},\cdots,t_{N}$).
\item The quantum network is represented by an undirected graph $G$. Each edge corresponds to a noiseless $q$-dimensional quantum channel, which can be used once per session in either direction. We assume that $q$ is a prime power.
\item Classical communication between any pair of nodes is freely available.
\item Terminal nodes may share a small amount of entanglement. The amount of shared entanglement is assumed to be negligible compared to the dimensions of the input quantum states.
\end{enumerate}

The fourth condition implies that terminal nodes may share entanglement, but the amount is insufficient to simulate a $q$-dimensional quantum communication channel via quantum teleportation. We also allow the graph $G$ to have multiple edges, all with the same weight $q$. Hence, the total quantum communication capacity between two nodes $u$ and $v$ is given by the number of edges connecting them.

Under this setting, suppose that $M_l$ copies of an unknown $d_l$-dimensional input state $\ket{\Psi_l} \in \mathbb{C}^{d_l}$ are available at each source node $s_l$, where $M_l \le N$ for all $l$. The goal of network coding is to construct a protocol that outputs an optimal symmetric universal clone of $\ket{\Psi_l}$ at every terminal node $t_n$ for all $l$ and $n$. We refer to such a protocol as multi-source multicast network coding for symmetric universal cloning of $d_l$-dimensional states with $M_l$ copies.

We derive the following results for this type of multi-source multicast network coding on the quantum network represented by $G$. We show that if there exists a classical linear multi-source multicast network code over an alphabet of size $q$ that achieves multicast rates $\lceil \log_q d_l[M_l] \rceil$ for all $l$ on an acyclic directed graph $G'$, then a corresponding quantum multi-source multicast network coding protocol for symmetric universal cloning can be constructed. Here, the directed graph $G'$ is obtained by assigning directions to the edges of $G$, and we assume that one of the terminal nodes shares $L \log_2 N$ ebits of entanglement with all other terminal nodes. The quantity $d[M] := \frac{(d+M-1)!}{M!(d-1)!}$ denotes the dimension of the symmetric subspace of $(\mathbb{C}^d)^{\otimes M}$. We note that the amount of entanglement required depends on $L$ and $N$, but not on the dimensions $d_l$ of the input states, and is therefore negligible when $d_l$ is large.

Further, suppose that there is only a single source node and that only a single copy of the input state is given at the source node.
The above result can then be rephrased as follows.
For a sufficiently large prime power $q$, single-source multicast network coding of a $q^r$-dimensional input state for symmetric universal cloning is possible on the quantum network $G$
with a small amount of entanglement shared among the terminal nodes,
provided that there exists an acyclic directed graph $G'$ 
obtained by assigning directions to the edges of $G$ and that the minimum cut between the source node and each terminal node is greater than or equal to $r$.
Therefore, in this case, symmetric universal clones can be multicast at a rate determined by the minimum cuts of the network. 

\begin{figure}
    \centering
\includegraphics[width=0.8\linewidth]{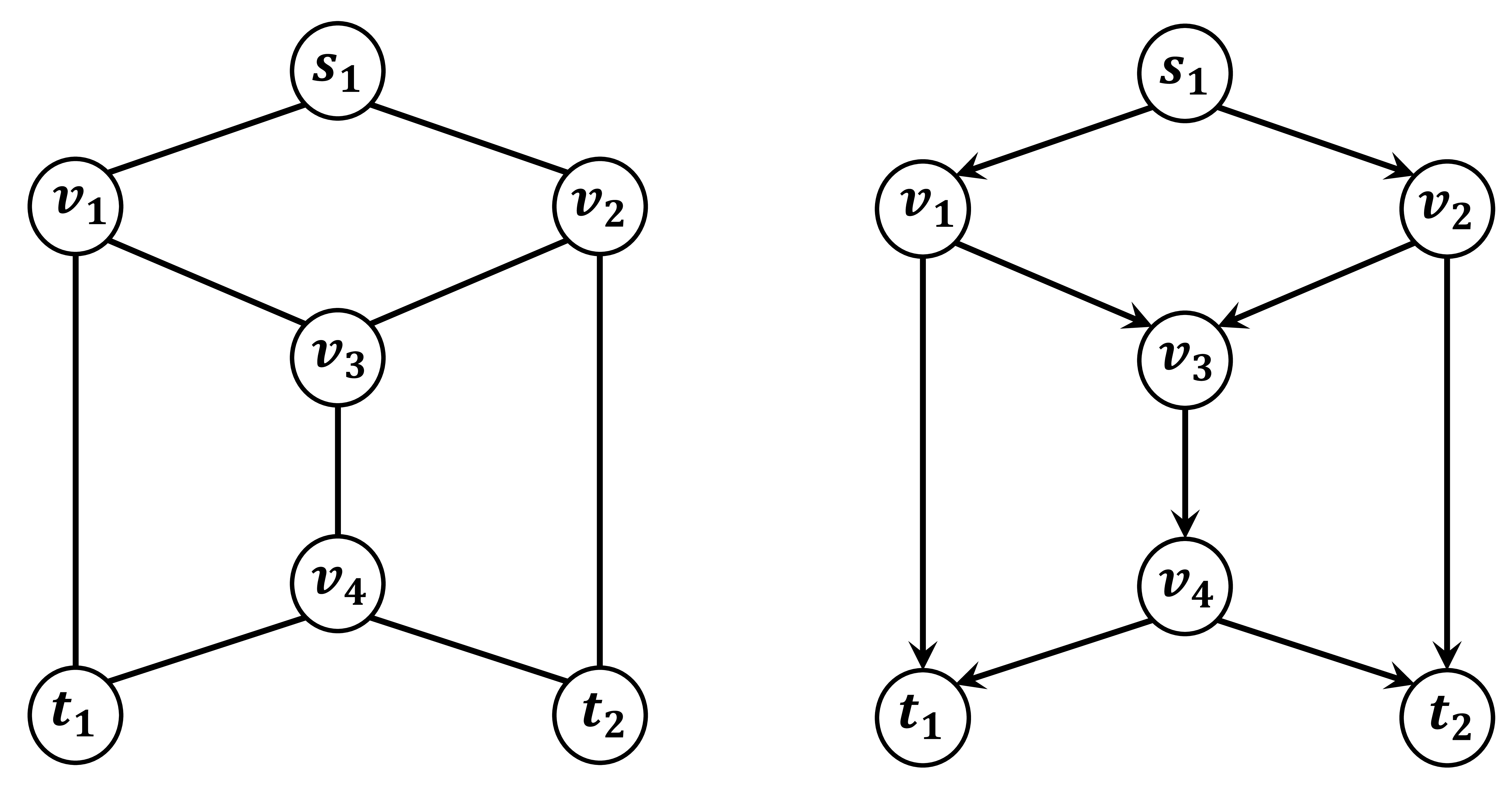}
    \caption{Undirected butterfly network (left) and directed butterfly network (right): These networks have one source node $s_1$, two terminal nodes $t_1$ and $t_2$, and three intermediate nodes $v_1$, $v_2$, and $v_3$.}
    \label{fig:butterfly1}
\end{figure}

Here, we provide a simple example to illustrate our results.
Suppose that $G$ is the undirected butterfly network shown on the left-hand side of Fig.~\ref{fig:butterfly1}, and that $G'$ is the corresponding directed butterfly network shown on the right-hand side of Fig.~\ref{fig:butterfly1}. Both $G$ and $G'$ have a single source node $s_1$ and two terminal nodes $t_1$ and $t_2$.
In this network, the minimum cut between $s_1$ and $t_1$, as well as the minimum cut between $s_1$ and $t_2$, is equal to $2$.
Hence, for a sufficiently large prime power $q$, multicast of an optimal symmetric single-source clone of a $q^2$-dimensional input state from $s_1$ to $t_1$ and $t_2$ is possible on $G$, with one ebit of entanglement shared between $t_1$ and $t_2$.
In fact, for butterfly networks, a classical multicast network code that achieves the minimum cut exists for any natural number $q \ge 2$ \cite{Yeung2007}. Therefore, the above result also holds for any natural number $q \ge 2$.

We also note that our main result holds in the special case $M_l=N$ for all $l$. 
In this situation, since the number of copies available at each source node coincides with the number of terminal nodes, symmetric universal cloning reduces to the identity operation. Consequently, the protocol simply distributes a single copy of the input state to all terminal nodes.
In the single-source setting, by taking into account the min-cut max-flow theorem, the main result can be reformulated in a manner similar to that described above. In this case, however, all minimum cuts must be at least $\lceil \log_q d[N] \rceil$ to distribute a $d$-dimensional input state. Generally, this throughput is greater than the throughput achieved using Shi et al.'s method \cite{Shi2006}.

We add an explanation regarding the limitations when practically using our protocol. Our problem setting assumes that all nodes are connected via noiseless quantum communication channels. It also assumes that error-free quantum operations are possible on each node. These assumptions are clearly not satisfied at the current technological level. Therefore, our protocol should be understood as one intended for use at a higher layer of future quantum networks where fault tolerant quantum computing and quantum repeater are implemented.

The foundational method underlying this work was previously disclosed in Japanese Patent JP6183958, which provides the protocol achieving our goal \cite{patent}. While the patent describes the protocol and basic analysis, detailed theoretical analysis has not been reported in the open literature.
Based on the patent, one of the authors constructed a similar type of protocol for optimal asymmetric universal cloning under the strong restriction that only a single copy of the input state is given, and there are two or three target nodes on the network \cite{app12126163}. 
Since asymmetric universal cloning is a generalization of symmetric universal cloning,
under the above strong restriction, the results of this paper are covered by the results in \cite{app12126163}.
On the other hand, this paper also covers the situation where there are more than three target nodes on the network, or multiple copies of input states are given in source nodes.

The rest of the paper is organized as follows:
We explain classical multicast network coding, Kobayashi et al.\'s quantum multicast network coding protocol, and symmetric universal cloning as preliminary knowledge in Section \ref{sec preliminary}. 
We give our protocol and the proof of the main result in Section \ref{sec main result}.  Finally, we give a summary in Section \ref{sec summary}.

\section{Preliminary}\label{sec preliminary}

In this section, as preliminary knowledge, we introduce a problem setting of classical multi-source
multicast network coding on Subsection \ref{subsec classical multicast}, Kobayashi et al.'s multicast quantum network coding protocol on Subsection \ref{subsec:Kobayashi-et-al.'s}, and symmetric optimal universal cloning on Subsection \ref{subsec cloning}. 

\subsection{Classical multi-source multicast network coding}\label{subsec classical multicast}

In this subsection, we present a problem setting of classical multicast network coding. In the classical setting of network coding, a network is given as a directed graph $G=(V,E)$. Here, a vertex $v\in V$
represents a node of the network, and an edge $e\in E$ represents a (noiseless) communication channel of the network. We permit multiple edges connecting the same pair of nodes. For simplicity, we assume \emph{$G$ is acyclic}. We denote the set of all outgoing edges of node $v$ as $O\left(v\right)$ and the set of all incoming edges of node $v$ as $I\left(v\right)$. When a single character in  $\mathbb{F}_{q}$ is transmitted from a vertex $u\in V$ to another vertex $v\in V$ via a channel, the channel is expressed as $(u,v)\in E$ in the directed
graph, where $\mathbb{F}_{q}$ is a finite field of order $q$. For simplicity,
we assume that \emph{$q$ is a prime power}. In one session of a
protocol, we assume that each edge $e\in E$ can send only a single
character in $\mathbb{F}_{q}$. 

Since our setting is multi-source multicast, there are $L$ source nodes $\left\{ s_l\right\} _{l=1}^{L}$, and $N$ terminal nodes $\left\{ t_n\right\} _{n=1}^{N}$. The purpose of this network is transmitting messages $\vec{x}_l\in\mathbb{F}_{q}^{r_l}$ from $s_l$ to all terminal nodes for all $l$, where messages $\vec{x}_l$ are independently chosen for all $l$, and $r_l$ is called the source rate. When $L=1$, the problem setting coincides with a multicast setting, where there is a single source node $s_{1}$ from which message $\vec{x}_{1}\in\mathbb{F}_{q}^{r_1}$ is transmitted to all terminal nodes. 
We denote the sets of source nodes and terminal nodes by $V_{S}$ and $V_{T}$, respectively. Here, for simplicity, we assume that the source node $s_l$ does not have any incoming edges. 
This assumption does not lose generality, since by adding fictitious source node $s_l'$ and $r_l$ edges from $s_l'$ to $s_l$ on the graph, we can always make $s_l$ an intermediate node, and the new source node $s_l'$ does not have any incoming edge. The network flow of this new graph and the original graph is essentially the same. By a similar discussion, we can assume that the terminal node $t_n$ does not have any outgoing edge. We will use this assumption about $G$ throughout the remainder of this paper.

A multi-source multicast network coding is the following protocol:
At the beginning of the protocol, for all $l$, message $\vec{x}_l\in\mathbb{F}_{q}^{r_l}$ is given at $s_l$, and $s_l$ transmits an information $y\left(e'\right)$ to $e'$ for all outgoing edges $e'\in O\left(s_l\right)$. Here, $y\left(e'\right)$ is produced by a (in general non-linear) function
$f_{e'}$ of $\vec{x}_l$ as $y\left(e'\right)=f_{e'}$$\left(\vec{x}_l\right)$,
where $f_{e'}:\ \mathbb{F}_{q}^{r_l}\rightarrow\mathbb{F}_{q}$.
Then, all the intermediate node $v\in V-V_{S}-V_{T}$ behaves as follow: as soon as information $y\left(e\right)\in\mathbb{F}_{q}$ is transmitted to $v$ from all incoming edges $e\in I\left(v\right)$ of $v$, $v$ transmits an information $y\left(e'\right)$ to $e'$ for all outgoing edges $e'\in O\left(v\right)$. Here, $y\left(e'\right)$ is produced by a (in general non-linear) function $f_{e'}$ of $\left\{ y\left(e\right)\right\} _{e\in I\left(v\right)}$
as $y\left(e'\right)=f_{e'}$$\left(\left\{ y\left(e\right)\right\} _{e\in I\left(v\right)}\right)$,
where $f_{e'}:\ \mathbb{F}_{q}^{\left|I\left(v\right)\right|}\rightarrow\mathbb{F}_{q}$.
Finally, a terminal node $t_n$ generates output information $\vec{z}_{ln}:=g_{ln}\left(\left\{ y\left(e\right)\right\} _{e\in I\left(t_n\right)}\right)$ for all $l$ and $n$, where $g_{ln}:\ \mathbb{F}_{q}^{\left|I\left(t_n\right)\right|}\rightarrow\mathbb{F}_{q}^{r_{l}}$.
When $\vec{x}_l=\vec{z}_{ln}$ for all $l,n$ and for all $\vec{x}_l\in\mathbb{F}_{q}^{r_l}$,
this multisource-multicast network coding problem with rate $r_l$ is called solvable. When $L=1$, the protocol is called a single-source multicast network coding.

When $L=1$, a single source case, for a given directed graph $G=\left(V,E\right)$, source node $s_{1}\in V$, and a set of terminal nodes $V_{T}\subset V$, a necessary and sufficient condition for existence of a multicast network code is known as the following form (called \emph{max-flow and min-cut theorem of multicast network coding} \cite{Ahlswede2000,Yeung2007}):
A multicast network code with source rate $r_{1}$ exists for sufficiently large $q$, if and
only if a minimum-cut between $s_{1}$ and $t_n$ is greater than
or equal to $r_{1}$ for all $n$. Here, the minimum cut between a source node $s$ and a terminal node $t$ is defined as the minimum of cut-capacity
$C\left(S,T\right)$ over all $s$-$t$ cut $(S,T)$, where $s\in S$,
$t\in T$, $S\cup T=V$, and $S\cap T=\emptyset$. For such $V$ and $S$, Cut-capacity $C\left(S,T\right)$
is defined as a number of edges $\left(u,v\right)\in E$ satisfying
$u\in S$ and $v\in T$. Hence, minimum-cut $X_{min}\left(s,t\right)$
between $s$ and $t$ can be written down as 
\begin{align}
X_{min}\left(s,t\right):= & min_{S,T\in2^{V}}\big\{ C\left(S,T\right)|\ s\in S,v\in T, \nonumber \\
 & \qquad S\cup T=V,\mbox{\ and}\ S\cap T=\emptyset\big\}, \label{eq def min cut}
\end{align}
and the above necessary and sufficient condition can be written down
as 
\[
X_{min}\left(s_{1},t_n \right)\ge r_{1}\quad\left(1\le\forall n\le N\right).
\]
Furthermore, it is known that to attain the above bound, all
functions $f_{v}$ and $g_{ln}$ can be chosen as linear functions \cite{Yeung2007}. 

\subsection{Kobayashi et al.'s quantum multicast network coding\label{subsec:Kobayashi-et-al.'s}}

In the problem setting of quantum network coding with free classical communication, a quantum network is given as an \emph{undirected} graph $G=(V,E)$.
Here, a vertex $v\in V$ represents a node (that is, a quantum computer)
and an edge $e\in E$ represents a (noiseless) quantum channel. We
again permit multiple edges connecting the same pair of nodes. 
In a single session of a protocol on the quantum network, we assume
that each edge $(u,v)\in E$ can send one $q$-dimensional quantum
system \emph{either from $u$ to $v$ or from $v$ to $u$}. Note
that the reason why undirected edges are used in this definition is
explained as follows: In the case when we permit free classical communication, even if we priorly determine the direction of all edges, a state can be transmitted in the inverse direction by means of quantum teleportation \cite{Leung2010}.
We assume that\emph{ $q$ is a prime power}. Each node $v\in V$ can implement all quantum operations, including measurements, and also send classical information (that is, measurement outcomes) to any other node, instantaneously. 

In multi-source multicast settings, there are $L$ source nodes $\left\{ s_l\right\} _{l=1}^{L}$,
and $N$ terminal nodes  $\left\{ t_n\right\}_{n=1}^{N}$. The purpose
of this quantum network may be transmitting unknown quantum states $\ket{\Psi_l}\in \mathcal{U}_l$ from $s_l$ to all terminal nodes for all $l$, where $\dim\mathcal{U}_l=q^{r_l}$
and $r_l$ is a source rate. However, because of the no-cloning theorem \cite{WZ82}, this task is clearly prohibited in quantum mechanics. Hence, the purpose of multi-source multicast QNC should be producing a $N$-partite clone $\rho_l$ of $\ket{\Psi_l}$ on system $\mathcal{V}_{l1}\otimes\cdots\otimes\mathcal{V}_{lN}$ for all $l$, where system $\mathcal{V}_{ln}$ is $q^{r_{l}}$-dimensional
space on node $t_n$. 

Kobayashi et al. chose a $N$-partite clone $\rho_l=:\ket{\Phi_l}\bra{\Phi_l}$
as a GHZ-type state \cite{Kobayashi2010}; that is, when $\ket{\Psi_l}:=\sum_{k\in\mathbb{F}_{q}^{r_l}}\alpha_{k}\ket{k}_{\mathcal{U}_l}$,
$\ket{\Phi_l}\in\bigotimes_{n=1}^{N}\mathcal{V}_{ln}$ is given
as follows: 
\[
\ket{\Phi_l}:=\sum_{k\in\mathbb{F}_{q}^{r_l}}\alpha_{k}\ket{k}_{\mathcal{V}_{l1}}\otimes\cdots\otimes\ket{k}_{\mathcal{V}_{lN}},
\]
 where $\left\{ \ket{k}_{\mathcal{H}}\right\} _{k\in\mathbb{F}_{q}^{r_l}}$
is a computational basis of $q^{r_l}$-dimensional space $\mathcal{H}$.
They further showed that if $\ket{\Phi_l}$ is produced among all terminal nodes $t_n$, then $\ket{\Psi_l}$ can be afterward derived on any terminal node $t_n$ anytime we want by local operations and classical communication (LOCC) \cite{Kobayashi2010}.
Hence, when $\ket{\Phi_l}$ can be produced on terminal nodes in the above problem setting, Kobayashi et al.'s multi-source multicast QNC is called solvable with rate $r_l$. Kobayashi et al. showed that their network coding problem on an acyclic undirected graph $G$ is solvable if classical linear multi-source multicast network coding is solvable with rate $r_l$ on a directed graph $G'$. Here, $G'$ is necessary to be derived from $G$ by adding directions of edges; that is, $G'$ is equal to $G$ except directions of edges. 

Before presenting Kobayashi et al.'s protocol, we need to define several
notations: For a function $f:\ \mathbb{F}_{q}^{r}\rightarrow\mathbb{F}_{q}$,
an controlled unitary operator $U_{f}$ on $\left(\mathbb{C}^{q}\right)^{\otimes r}\otimes\mathbb{C}^{q}$
is defined by
\begin{align*}
U_{f}:=\sum_{y_{1,}\cdots,y_{r+1}\in\mathbb{F}_{q}} & \ket{y_{1},\cdots,y_{r}}\bra{y_{1},\cdots,y_{r}}\\
 & \otimes\ket{y_{r+1}+f\left(y_{1,}\cdots,y_{r}\right)}\bra{y_{r+1}}.
\end{align*}

Suppose a prime power can be written as $q=p^h$ for a prime number $p$. Then, for $\beta \in \mathbb{F}_q$, generalized Pauli operation $Z_\beta$ on $\mathcal{H}\simeq\mathbb{C}^{q}$
is defined by 
\[
Z_\beta:=\sum_{k\in\mathbb{F}_{q}}\omega^{\mathrm{tr}k\beta}\ket{k}\bra{k},
\]
where $\omega:=e^{2\pi i/p}$ is a $p$-th root of unity. 
Here, for $z\in \mathbb{F}_q$, $\mathrm{tr}z$ is defined by $\mathrm{Tr}\psi (z)$, where $\psi (z)$ is the matrix representation of the multiplication map $x \mapsto zx$. 
We note that $\mathbb{F}_q$ can be identified with the vector space $\mathbb{F}_p^h$.
In a space $\mathcal{H}\simeq\mathbb{C}^{q}$, the Fourier basis $\left\{ \ket{\tilde{k}}\right\} _{k\in\mathbb{F}_{q}}$
with respect to a computational basis $\left\{ \ket{k}\right\} _{k\in\mathbb{F}_{q}}$
is defined by 
\[
\ket{\tilde{k}}:=q^{-1/2}\sum_{l\in\mathbb{F}_{q}}\omega^{\mathrm{tr}kl}\ket{l}
\]

Suppose classical multi-source multicast QNC on $\mathbb{F}_q$ is solvable with rate $r_l$ on an acyclic directed graph $G'$. Then, there exist linear functions $f_{e}$ and $g_{ln}$ achieving this task. For $e\in E\left(G'\right)$,
we denote its destination $v_{d}\left(e\right)$ and its origin $v_{o}\left(e\right)$.
The set of all outgoing edges of $v$ is written as $O\left(v\right)$, and the set of all incoming edges of $v$ is written as $I\left(v\right)$.
By using these functions, Kobayashi et al's protocol is given as follows:
\begin{enumerate}
\item At the beginning of the protocol, source node $s_l$ prepares $q^{r_l}$-dimensional
system $\mathcal{U}_l$ for all $l$. System $\otimes_{l=1}^{L}\mathcal{U}_l$ is initialized in an arbitrary input state. For all $e\in E\left(G'\right)$, node $v_{o}\left(e\right)$ prepares $q$-dimensional system $\mathcal{H}\left(e\right)$,
and initializes it in $\ket{0}$. Terminal node $t_n$ prepares
$q^{r_l}$-dimensional system $\mathcal{V}_{ln}$ for all $l$ and
$n$, and initializes it in $\ket{0}$.
\item For all $l$, source node $s_l$ behaves as follows: $s_l$ implements
$U_{f_{e}}$ on system $\mathcal{U}_l\otimes\mathcal{H}\left(e\right)$
for all $e\in O\left(s_l\right)$. Then, $s_l$ measures $\mathcal{U}_l$
in the Fourier basis, and derives a classical outcome $\xi_l$.
Then, $s_l$ sends $\xi_l$ to all terminal nodes $t_n$, and also transmits $\mathcal{H}\left(e\right)$ to node $v_{d}\left(e\right)$ for all $e\in O\left(s_l\right)$. 
\item All intermediate nodes $v\in V-V_{S}-V_{T}$ behave as follows: as
soon as all systems $\mathcal{H}\left(e'\right)$ satisfying $e'\in I\left(v\right)$
arrive at $v$, $v$ first implements $U_{f_{e}}$ on system $\left(\bigotimes_{e'\in I\left(v\right)}\mathcal{H}\left(e'\right)\right)\otimes\mathcal{H}\left(e\right)$
for all $e\in O\left(v\right)$. Second, $v$ measures $\mathcal{H}\left(e'\right)$
in the Fourier basis, and derives a classical outcome $\eta\left(e'\right)$
for all $e'\in I\left(v\right)$. Third, $v$ sends $\eta\left(e'\right)$
to all terminal nodes $t_n$ for all $e'\in I\left(v\right)$,
and also transmits $\mathcal{H}\left(e\right)$ to node $v_{d}\left(e\right)$
for all $e\in O\left(v\right)$.
\item All terminal nodes $t_n$ behave as follows: as soon as all systems
$\mathcal{H}\left(e'\right)$ satisfying $e'\in I\left(t_n\right)$
arrive at $t_n$, $t_n$ implements $U_{g_{ln}}$ on system $\left(\bigotimes_{e'\in I\left(t_n\right)}\mathcal{H}\left(e'\right)\right)\otimes\mathcal{V}_{ln}$
for all $l$ and $n$. $t_n$ measures $\mathcal{H}\left(e'\right)$
in the Fourier basis, and derives a classical outcome $\eta\left(e'\right)$
for all $e'\in I\left(t_n\right)$. 
\item Finally, to fix errors induced by measurements, depending
on $\left\{ \xi_l\right\} _{l=1}^{L}$ and $\left\{ \eta\left(e'\right)\right\} _{e'\in E}$,
$t_n$ implements generalized Pauli $Z_{\beta_{ln}}$ onto system $\mathcal{V}_{ln}$
with appropriately chosen $\beta_{ln}\in \mathbb{F}_q$.
\end{enumerate}
After these steps, system $\bigotimes_{ln}\mathcal{V}_{ln}$ can be
found in the desire output state. In the above step 5, $\beta_{ln}$ must be determined by $f_{e}$, $g_{ln}$,
$\left\{ \xi_l\right\} _{l=1}^{L}$ and $\left\{ \eta\left(e'\right)\right\} _{e'\in E}$ in a slightly complicated way. We do not explain the details in this paper.
For the details, please see \cite{Kobayashi2010} and \cite{9427210}. 

When $L=1$ (a single source case), by using the max-flow and min-cut theorem of a single source multicast classical network coding, Kobayshi et al's result can be rephased as follows: \emph{For sufficiently large $\mathbb{F}_q$, Kobayashi et al.'s multicast QNC on an undirected graph $G$ is solvable with source rate $r$, if there exists a directed graph $G$ such that $G'$ can be derived from $G$ by adding directions of edges, and min-cut $X_{min}(s_1,t_n)$
between source node $s_{1}$ and terminal node $t_n$ is greater than and equal to $r$ for all $n$.}
Here, $X_{min}(s,t)$ is defined by Eq.(\ref{eq def min cut})

\subsection{Optimal univeral symmetric cloning}\label{subsec cloning}

In this subsection, we introduce known results and notations about optimal symmetric universal cloning, which is necessary to present our multi-source multicast quantum network coding in the next section.
Various different schemes for approximate quantum cloning are known \cite{RevModPhys.77.1225,FAN2014241}.
Among them, in this paper, we focus on an optimal symmetric universal
quantum cloning machine (UQCM) \cite{BH1996,GM1997,BH1998,Werner1998}.
Here, ``\emph{universal}'' means that fidelity between a clone and an input state does not depend on the input state. ``\emph{Symmetric}'' means that fidelity between a clone and the input state does not depend on the choice of a clone. Hence, an optimal symmetric UQCM is the best cloning protocol having these properties.

Suppose $M$-copies of state $\ket{\psi}\in\mathbb{C}^{d}$ are given, and we try to make $N$-clones of $\ket{\psi}$. An optimal symmetric UQCM in this situation (called $M\rightarrow N$ optimal symmetric UQCM) is given by
\begin{align}
  \Lambda_{UQCM}\left(P\left(\ket{\psi}^{\otimes M}\right)\right)
:=  \frac{d\left[M\right]}{d\left[N\right]}P_{{\rm sym}}^{\left(N\right)}\left(P(\ket{\psi}^{\otimes M})\otimes I_d^{\otimes N-M}\right)P_{{\rm sym}}^{(N)},\label{eq:def uqcm}
\end{align}
 where $P\left(\ket{\psi}\right)$ is defined by $P\left(\ket{\psi}\right):=\ket{\psi}\bra{\psi}$,
$P_{{\rm sym}}^{\left(N\right)}$ is a projection onto the symmetric
subspace of $\left(\mathbb{C}^{d}\right)^{\otimes N}$, $I_d$ is an identity operator on $\mathbb{C}^{d}$, and $d\left[N\right]$ is defined as $d\left[N\right]:=\mathrm{Tr} P_{sym}^{\left(N\right)}=\frac{\left(d+N-1\right)!}{N!\left(d-1\right)!}$.
We will write the symmetric subspace of $\left(\mathbb{C}^{d}\right)^{\otimes N}$
as $Sym\left(\mathbb{C}^{d}\right)^{\otimes N}$.

In the remaining part of this subsection, we introduce another formula
which is equivalent to Eq.\eqref{eq:def uqcm}, which plays an important
role in the next section. First, to make equations appearing in the remaining part of the paper simpler, we define a computational basis
of an input state as $\left\{ \ket{x}\right\} _{x\in\mathbb{Z}_{d}}$,
where $\mathbb{Z}_{d}$ is a cyclic group defined by  $\mathbb{Z}_{d}:=\mathbb{Z}/d\mathbb{Z}=\left\{ 0,\cdots,d-1 \right\} $.
Then, an input state can be written
as 
\begin{eqnarray*}
\ket{\psi} & = & \sum_{x\in\mathbb{Z}_d}\alpha_{x}\ket{x}
\end{eqnarray*}
by using complex coefficients $\alpha_{x}\in\mathbb{C}$. We further
introduce necessary notations: $\mathbb{Z}_{d}^N$ is a set of lists, or an abelian group, defined as $\mathbb{Z}_{d}^N:=\overbrace{\mathbb{Z}_{d}\times\cdots\times\mathbb{Z}_{d}}^{N}$.
Then, $\{\ket{\vec{z}}\}_{\vec{z}\in\mathbb{Z}_{d}^{N}}$ can be considered as an orthonormal basis on $(\mathbb{C}^d)^{\otimes N}$. $\mathbb{Z}_{d}^{+N}$ is defined as a subset of $\mathbb{Z}_{d}^{N}$ such that any element $\vec{x}\in\mathbb{Z}_{d}^{+N}$ satisfies $x_{j}\le x_{j+1}$ for
all $j$, where $x_{j}$ is $j$th entry of $\vec{x}$:$ $ 
\[
\mathbb{Z}_{d}^{+N}:=\left\{ \vec{x}\in\mathbb{Z}_{d}^{N}\ |\ x_{j}\le x_{j+1}\ \left(\forall j\right)\right\} .
\]
For $\vec{z}\in\mathbb{Z}_{d}^{+N}$, $\ket{\vec{z}}_{{\rm sym}}\in(\mathbb{C}^d)^{\otimes N}$
is defined as 
\begin{align}
\ket{\vec{z}}_{{\rm sym}}:=|\Omega_{\vec{z}}|^{1/2}\cdot P_{sym}^{\left(N\right)}\ket{\vec{z}} & =\frac{|\Omega_{\vec{z}}|^{1/2}}{N!}\sum_{\sigma\in S_N}\ket{\sigma\vec{z}}\nonumber \\
 & =\frac{1}{|\Omega_{\vec{z}}|^{1/2}}\sum_{\vec{z}'\in\Omega_{\vec{z}}}\ket{\vec{z}'},\label{eq:def ket vec z sym}
\end{align}
 where $S_{N}$ is the symmetric group of order $N$, and $\Omega_{\vec{z}}$ is a set of all permuted list of $\vec{z}$ defined by 
\begin{equation}\label{eq def Omega}
\Omega_{\vec{z}}:=\left\{ \sigma\vec{z}\ |\ \sigma\in S_{N}\right\}.
\end{equation}
Here, for $\sigma \in S_N$ and $\vec{z}=(z_1,\cdots, z_N)$, $\sigma \vec{z}$ is defined by
\begin{equation}\label{eq def sigma vec z}
\sigma \vec{z}:=(z_{\sigma(1)},\cdots, z_{\sigma(N)}).    
\end{equation}
To derive the last equality of Eq.(\ref{eq:def ket vec z sym}), 
we use 
\begin{equation}\label{eq sum ket sigma vec z H vec z}
    \sum_{\sigma \in S_N}\ket{\sigma \vec{z}} = |H_{\vec{z}}|\sum_{\vec{z}\in \Omega_{\vec{z}}}\ket{\vec{z}},
\end{equation}
where $H_{\vec{z}}$ is a stabilizer of $\vec{z}$ defined by 
\begin{equation}
    H_{\vec{z}}:=\{\sigma \in S_N \ |\ \sigma \vec{z}=\vec{z} \}. 
\end{equation}
The proof of Eq.(\ref{eq sum ket sigma vec z H vec z}) is given on Appendix.A.
Eq.(\ref{eq:def ket vec z sym}) leads that $\{\ket{\vec{z}}_{sym}\}_{\vec{z}\in\mathbb{Z}_{d}^{+N}}$ forms
an orthonormal basis of the symmetric subspace of $(\mathbb{C}^{d})^{\otimes N}$.

By using $I_d=\sum_{x\in\mathbb{Z}_{d}}\ket x\bra x$, the right-hand
side of Eq.\eqref{eq:def uqcm} can be rewritten as follows: 
\begin{eqnarray}
&\quad & \Lambda_{UQCM}\left(P\left(\ket{\psi}^{\otimes M}\right)\right)\nonumber \\
& := & \frac{d\left[M\right]}{d\left[N\right]}\sum_{\vec{y}\in\mathbb{Z}_{d}^{N-M}}P\left(P_{{\rm sym}}^{\left(N\right)}\ket{\psi}^{\otimes M}\ket{\vec{y}}\right)\nonumber \\
 & = & \frac{d\left[M\right]}{d\left[N\right]}\sum_{\vec{y}\in\mathbb{Z}_{d}^{+N-M}}|\Omega_{\vec{y}}|P\left(P_{{\rm sym}}^{\left(N\right)}\ket{\psi}^{\otimes M}\ket{\vec{y}}\right)\nonumber \\
 & = & \sum_{\vec{y}\in\mathbb{Z}_{d}^{+N-M}}P\Big(\sum_{\vec{x}\in\mathbb{Z}_{d}^{M}}\left(\prod_{j=1}^{M}\alpha_{x_{j}}\right)\cdot\sqrt{\frac{d\left[M\right]}{d\left[N\right]}\frac{|\Omega_{\vec{y}}|}{|\Omega_{\vec{x}\vec{y}}|}}\ket{\vec{x}\vec{y}}_{{\rm sym}}\Big),\label{eq:uqcm_intermediate}
\end{eqnarray}
where $\vec{x}\vec{y}:=\left(x_{1},\cdots,x_{M},y_{1},\cdots,y_{N-M}\right)\in\mathbb{Z}_{N}$
is a conjunction of lists $\vec{x}$ and $\vec{y}$, and we used 
\[
P_{{\rm sym}}^{\left(N\right)}\ket{\vec{x}}\ket{\vec{y}}=\frac{1}{\left|\Omega_{\vec{x}\vec{y}}\right|^{1/2}}\ket{\vec{x}\vec{y}}_{sym}
\]
in the third equality. We next define an isometry $V_{\vec{y}}:Sym\left(\mathbb{C}^{d}\right)^{M}\rightarrow Sym\left(\mathbb{C}^{d}\right)^{N}$
for a given  $\vec{y}\in\mathbb{Z}_{d}^{N-M}$ by 
\begin{eqnarray}
V_{\vec{y}}\ket{\vec{x}}_{{\rm sym}}=\ket{\vec{x}\vec{y}}_{{\rm sym}}.\label{eq:def_isometry_V}
\end{eqnarray}
It is convenient to extend the domain of $V_{\vec{y}}$ to $\left(\mathbb{C}^{d}\right)^{M}$
by defining $V_{\vec{y}}\ket{\phi}=0$ for all $\ket{\phi}$ on the
orthogonal complement of $Sym\left(\mathbb{C}^{d}\right)^{M}$. Further,
we define a Hermitian operator $K_{\vec{y}}$ on $Sym\left(\mathbb{C}^{d}\right)^{M}$ for $\vec{y}\in\mathbb{Z}_{d}^{+N-M}$
by 
\begin{equation}
K_{\vec{y}}\ket{\vec{x}}_{sym}=\sqrt{\frac{d\left[M\right]}{d\left[N\right]}\frac{|\Omega_{\vec{y}}|}{|\Omega_{\vec{x}\vec{y}}|}}\ket{\vec{x}}_{sym}.\label{eq:def_POVM_K}
\end{equation}
Again, we extend the domain of $K_{\vec{y}}$ to $\left(\mathbb{C}^{d}\right)^{M}$
by defining $K_{\vec{y}}\ket{\phi}=0$ for all $\ket{\phi}$ on the
orthogonal complement of $Sym\left(\mathbb{C}^{d}\right)^{M}$. Then,
from Eq.\eqref{eq:def_POVM_K}, we can check that $\{K_{\vec{y}}^{2}\}_{\vec{y}\in\mathbb{Z}_{d}^{+N-M}}$
is POVM on $Sym\left(\mathbb{C}^{d}\right)^{M}$ :
\[
\sum_{\vec{y}\in\mathbb{Z}_{d}^{+N-M}}K_{\vec{y}}^{2}=P_{{\rm sym}}^{\left(M\right)}.
\]
This fact can be shown by using the following equation:
\[
d\left[N\right]=\sum_{\vec{y}\in\mathbb{Z}_{d}^{+N-M}}\frac{d\left[M\right]|\Omega_{\vec{y}}|}{|\Omega_{\vec{x}\vec{y}}|}.
\]
 The above equation can be derived from the fact that $Tr\Lambda_{UQCM}\left(\ket{\vec{x}}\bra{\vec{x}}\right)=1$
for all $\vec{x}\in\mathbb{Z}_{d}^{M}$. 

Now, we are ready to present the main statement on this subsection.
Eqs.\eqref{eq:def ket vec z sym}, \eqref{eq:uqcm_intermediate}, \eqref{eq:def_isometry_V},
and \eqref{eq:def_POVM_K} lead the following formula:
\begin{align}
  \Lambda_{UQCM}\left(P\left(\ket{\psi}^{\otimes M}\right)\right)& =  \sum_{\vec{y}\in\mathbb{Z}_{d}^{+N-M}}V_{\vec{y}}K_{\vec{y}}P\left(\ket{\psi}^{\otimes M}\right)K_{\vec{y}}^\dagger V_{\vec{y}}^\dagger
  \nonumber \\
&=  \sum_{\vec{y}\in\mathbb{Z}_{d}^{+N-M}}
P\big(V_{\vec{y}}K_{\vec{y}}\ket{\psi}^{\otimes M} \big).\label{eq:uqcm_K_V}
\end{align}
 This equation shows that we can implement a symmetric UQCM on $\ket{\psi}^{\otimes M}\in Sym\left(\mathbb{C}^{d}\right)^{M}$
by first applying measurement $\{K_{\vec{y}}\}_{\vec{y}\in\mathbb{Z}_{d}^{+N-M}}$ and then, isometry $V_{\vec{y}}$. 
We will use this fact to construct our multi-source multicast QNC protocol in the next section.

\section{Multi-source multicast QNC by UQCM}\label{sec main result}

In this section, we present the main results of this paper,
that is, a protocol which multicasts a quantum state over a quantum
network in the form of optimal symmetric universal cloning.
In this section, we denote $\{1,\cdots, n\}$ as $[n]$. 

\subsection{Problem setting and main theorem}\label{subsec main theorem}
In this subsection, we introduce the problem setting of multi-source multicast QNC for symmetric universal cloning, and the main theorem of this paper.

A mathematical model of a quantum network is mostly the same
as the one explained in Section \ref{subsec:Kobayashi-et-al.'s},
which is used in Kobayashi et al.'s protocol. The only difference is
that we permit terminal nodes to share a small amount of entanglement in our problem setting.
In our settings, there are $L$ source nodes $\left\{ s_{l}\right\} _{l=1}^{L}$, and $N$ terminal nodes $\left\{ t_{n}\right\} _{n=1}^{N}$ on a quantum network $G=(V,E)$, where each quantum channel $e\in V$ can send a $q$-dimensional system with prime power $q$, and free classical communication is available. We assume that $M_{l}$ copies of $d_l$-dimensional states   $\ket{\Psi_{l}}\in\mathbb{C}^{d_{l}}$
is given on each source node $s_{l}$, where $M_l$ satisfies $M_l \le N$ for all $l\in [L]$.
The goal of our multi-source multicast
QNC, which is different from Kobayashi et al.'s protocol, is producing
a $N$-partite \emph{optimal symmetric universal clone} $\rho_{l}$ of $\ket{\Psi_{l}}$ on system $\mathcal{V}_{l1}\otimes\cdots\otimes\mathcal{V}_{lN}$ for all $l\in [L]$, where system $\mathcal{V}_{ln}$ is $d_l$-dimensional
space on node $t_{n}$. 
To achieve this goal, we further assume that $N$ terminal nodes $\left\{ t_{n}\right\} _{n=1}^{N}$ share an entanglement resource whose amount does not scale with respect to $d_l$.
In this paper, we use {\it ebit} as the unit of amount of entanglement: 
For example, the amount of entanglement of a maximally entangled state $(d)^{-1/2}\sum_{i=0}^{d-1}\ket{i}\otimes \ket{i}$ is $log_2 d$ ebit. 
Under these assumptions, if the above task is possible, we say ``{\it multi-source multicast QNC problem on $G$ for symmetric universal cloning with $L$ source nodes, $N$ terminal nodes and  $M_l$ copies of $d_l$-dimensional system for all $l\in [L]$ is solvable}''.

The main theorem of the paper is written as follows:\\
{\it Theorem: multi-source multicast QNC problem on $G$ for symmetric universal cloning  with $L$ source nodes, $N$ terminal nodes and $M_l$ copies of $d_l$-dimensional system  for all $l\in [L]$ is solvable, if 
one of the terminal nodes shares $L\log _2 N$ ebit with all other terminal nodes, and there exists an acyclic directed graph $G'$ such that $G'$ is derived by adding directions on $G$, and 
multi-source multicast classical network coding problem on acyclic directed graph $G'$ is solvable with rate $r_l=\lceil log_q  d[M_l] \rceil$. }
Here, $d[M]$ is defined by $d[M]=\frac{(d+M-1)!}{M!(d-1)!}$.

Suppose there is only single source node $s$ and only single copy of input state $\ket{\Psi}$ is given on $s$;  that is, $L=1$ and $M_1=1$.
In this case, by means of max-flow
and min-cut theorem of classical multicast network coding \cite{Ahlswede2000,Yeung2007},
the main theorem can be rephrased as follows:\\
{\it Corollary: for sufficiently large prime power $q$, single-source multicast QNC problem on $G$ for symmetric universal cloning  with terminal nodes $t_1,\cdots t_N$ and single copy of $q^r$-dimensional system is solvable, if 
one of the terminal nodes shares $L\log _2 N$ ebit with all other terminal nodes, and there exists an acyclic directed graph $G'$ such that $G'$ is derived by adding directions on $G$, and 
minimum cut $X_{min}(s,t_n)$ between $s$ and terminal node $t_n$ is greater than or equal to $r$ for all $n\in[N]
$. }
Here, $X_{min}(s,t_n)$ is defined by Eq.(\ref{eq def min cut}).

We clarify the assumption about the entanglement resource. 
Suppose $\mathcal{D}_1^{(l)}$ is $N!$-dimensional system on terminal node $t_1$ for all $l\in [L]$, 
and $\mathcal{C}_n^{(l)}$ is $N$-dimensional system on terminal node $t_n$ for all $n\in [N]$ and 
 $l\in  [L]$.
 The entanglement resource which is necessary for our protocol is $L$-copies of  $\ket{\Upsilon}$.
Here, $l$-th copies of $\ket{\Upsilon}$ is on the system $\mathcal{D}_1^{(l)}\otimes \big( \bigotimes _{n=1}^N \mathcal{C}_n^{(l)}\big)$ for all $l\in [L]$, and defined by
\begin{equation}\label{eq def Upsilon}
    \ket{\Upsilon}=\frac{1}{\sqrt{N!}}\sum_{\sigma \in S_N} \ket{\sigma}_{\mathcal{D}_1^{(l)}}\otimes \big( \bigotimes _{n=1}^N \ket{\sigma(n)}_{\mathcal{C}_n^{(l)}} \big).
\end{equation}
It is clear that 
if terminal node $t_1$ shares $L\log _2 N$ ebit with terminal nodes $t_n$ for all $2\le n \le N$, then, it is possible to share $L$-copies of $\ket{\Upsilon}$ among the terminal nodes.

$\ket{\Upsilon}\in \mathcal{D}_1^{(l)}\otimes \big( \bigotimes _{n=1}^N \mathcal{C}_n^{(l)}\big)$ defined by Eq.(\ref{eq def Upsilon}) is asymmetric with respect to 
$n$. That is, terminal node $t_1$ shares entanglement with $t_n$ for all $n\ge 2$, but terminal node $t_n$ does not share any entanglement with $t_{n'}$ for all $n\neq n'$ if $n,n' \ge 2$.  Thus, terminal node $t_1$ plays a special role in our protocol when $\ket{\Upsilon}$ is shared in this way. 
On the other hand, as we will see later, any other node can also play this special role, 
if it shares $L\log _2 N$ ebit with all other terminal nodes.
This is because, in this case, this special node can share $\ket{\Upsilon}$ with all other nodes so that the special node has system $\mathcal{D}_1^{(l)}\otimes \mathcal{C}_1^{(l)}$.

\subsection{Description of the protocol}\label{subsec protocol}
In this subsection, we introduce the protocol of multi-source multicast QNC as symmetric universal cloning. The problem setting is given in the previous subsection.

Before introducing the protocol, we first introduce the subroutine implemented by LOCC used in the protocol. 
Suppose there are three systems $\mathcal{A}$, $\mathcal{B}$ and $\mathcal{C}$; 
$\{ \ket{u}\}_{u=0}^{\dim \mathcal{A}-1}$ and $\{ \ket{v}\}_{v=0}^{\dim \mathcal{C}-1}$ are 
orthonormal basis on $\mathcal{A}$ and $\mathcal{C}$, respectively.
The purpose of the subroutine is transform the state $\ket{\Gamma} \in \mathcal{A}\otimes \mathcal{B} \otimes \mathcal{C}$ to the state
$\ket{\Gamma'} \in \mathcal{A}\otimes \mathcal{B}$ by LOCC,
where $\ket{\Gamma}$ and $\ket{\Gamma '}$ are defined by 
\begin{align}
\ket{\Gamma}&:=  \sum_{u=0}^{\mathcal{A}-1} \alpha_u \ket{u}_{\mathcal{A}}\ket{\phi_u}_{\mathcal{B}}\ket{f(u)}_{\mathcal{C}},\nonumber \\
\ket{\Gamma '}&:=  \sum_{u=0}^{\mathcal{A}-1} \alpha_u \ket{u}_{\mathcal{A}}\ket{\phi_u}_{\mathcal{B}},
\end{align}
respectively.
Here, $\{ \alpha_u\}_{u=0}^{\dim \mathcal{A}-1}$ is an arbitrary coefficient, $f$ is an arbitrary function from $\{0, \cdots , \dim \mathcal{A} -1 \}$ to
$\{0, \cdots , \dim \mathcal{C} -1 \}$, and $\ket{\phi_u}\in \mathcal{B}$ is an arbitrary state depending on $u$.  
We further define the Fourier basis $\{\ket{\tilde{p}}\}_{p=0}^{\dim \mathcal{C}-1}$ of $\{ \ket{v}\}_{v=0}^{\dim \mathcal{C}-1}$ by 
\begin{equation}
    \ket{\tilde{p}}:=\frac{1}{\sqrt{\dim \mathcal{C}}}\sum_{v=0}^{\dim \mathcal{C}-1}
    e^{\frac{pv}{d}2\pi i}\ket{v},
\end{equation}
and a unitary operator $U_p$ depending on $p\in \{0,\cdots, \dim \mathcal{C}-1\}$
by 
\begin{equation}
    U_p:= \sum_{u=0}^{\dim \mathcal{A}-1} e^{\frac{pf(u)}{d}2\pi i}\ket{u}\bra{u}.
\end{equation}
Then, the LOCC transformation from $\ket{\Gamma}$ to $\ket{\Gamma '}$ 
can be implemented by the following subroutine:\\
\begin{enumerate}
    \item $\mathcal{C}$ is measured in the Fourier basis $\ket{\tilde{p}}$, and the measurement outcome $p'$ is sent to the system $\mathcal{A}$.
    \item Depending on $p'$, a unitary $U_{p'}$ is applied on $\mathcal{A}$. 
\end{enumerate}
If $\mathcal{A}\otimes \mathcal{B} \otimes \mathcal{C}$ is on $\ket{\Gamma}$,
then, after the first step  $\mathcal{A}\otimes \mathcal{B}$ is on 
\begin{align}
\sum_{u=0}^{\mathcal{A}-1} e^{-\frac{p'f(u)}{d}i\pi}\alpha_u \ket{u}_{\mathcal{A}}\ket{\phi_u}.
\end{align}
It is clear that $\ket{\Gamma '}$ is derived by applying $U_{p'}$ on $\mathcal{A}$.
The application of this LOCC transformation will be called ``{\it removing $\mathcal{C}$ by using $\mathcal{A}$} with the LOCC subroutine''. 

We also introduce unitary operators used in our protocol. 
For all $l\in [L]$, a unitary operator $V_0^{(l)}$ on $\mathcal{A}_1\otimes \mathcal{G}_1\otimes\mathcal{D}\otimes\mathcal{F}$ is defined by 
\begin{align}\label{eq def V 0 l}
    &V_0^{(l)}\nonumber \\
    :=&\sum_{\vec{a}\in \mathbb{Z}_{d_l}^{+M_l}, \vec{x}\in \mathbb{Z}_{d_l}^{N-M_l},\sigma \in S_N, \vec{z}\in \mathbb{Z}_{d_l}^{N}}\ket{\vec{a},\vec{x},\sigma, \vec{z}+\sigma (\vec{a}\vec{x})}_{\mathcal{A}_1,\mathcal{G}_1,\mathcal{D},\mathcal{F}}
    \bra{\vec{a},\vec{x},\sigma, \vec{z}}_{\mathcal{A}_1,\mathcal{G}_1,\mathcal{D},\mathcal{F}},
\end{align}
where the dimensions of $\mathcal{A}_1,\mathcal{G}_1,\mathcal{D}_1,\mathcal{F}_1$ 
are $d_l[M_l]$, $d_l^{N-M_l}$, $N!$ and $d_l^N$, respectively.
For all $l\in [L]$ and $n\in [N]$, a unitary operator $V_n^{(l)}$ on $\mathcal{A}_n\otimes \mathcal{G}_n\otimes\mathcal{C}_n \otimes \mathcal{E}_n$ is defined by 
\begin{align}\label{eq def V n l}
   & V_n^{(l)}\nonumber \\
   :=& \sum_{\vec{a}\in \mathbb{Z}_{d_l}^{+M_l}, \vec{x}\in \mathbb{Z}_{d_l}^{N-M_l},n' \in [N], z\in \mathbb{Z}_{d_l}}\ket{\vec{a},\vec{x},n', z+(\vec{a}\vec{x})_{n'}}_{\mathcal{A}_n,\mathcal{G}_n,\mathcal{C}_n,\mathcal{E}_n}\nonumber \\
& \qquad \qquad \qquad \qquad \qquad \qquad \qquad \qquad \qquad  \qquad\bra{\vec{a},\vec{x},n', z}_{\mathcal{A}_n,\mathcal{G}_n,\mathcal{C}_n,\mathcal{E}_n},
\end{align}
where $(\vec{a}\vec{x})_{n'}$ is $n'$-th entry of $\vec{a}\vec{x}$, and the dimensions of ${\mathcal{A}_n,\mathcal{G}_n,\mathcal{C}_n,\mathcal{E}_n}$  
are $d_l[M_l]$, $d_l^{N-M_l}$, $N$ and $d_l$, respectively.

We introduce the protocol of  multi-source multicast QNC on $G$ with source nodes $s_1,\cdots, s_L$ and terminal nodes $t_1, \cdots, t_N$ for symmetric universal cloning  as follows:
\begin{description}
    \item[Step 0] Quantum systems are prepared on source nodes and terminal nodes as follows:
    $d_l^{M_l}$-dimensional systems $\mathcal{S}_l$ are on source node $s_l$ for all $l\in [L]$, respectively. $d_l[M_l]$-dimensional systems $\mathcal{A}_n^{(l)}$, $N$-dimensional systems $\mathcal{C}_n^{(l)}$, $d_l$-dimensional systems $\mathcal{E}_n^{(l)}$, 
    $d_l^{N-M_l}$-dimensional systems $\mathcal{G}_n^{(l)}$ are 
    on terminal node $t_n$ for all $n\in [M]$ and $l\in [L]$.
        $N!$-dimensional systems $\mathcal{D}^{(l)}$ and $d_l ^N$-dimensional systems $\mathcal{F}^{(l)}$ are on terminal node $t_1$ for all $l\in [L]$. 
     
    \item[Step1] The initial states $\ket{\Psi_l}^{\otimes M_l}$ are given on $\mathcal{S}_l$ for all $l \in [L]$. For all $l\in [L]$, source node $s_l$ measures $\mathcal{S}_l$ by measurement $\{K_{\vec{y}_l}\}_{\vec{y}_l\in\mathbb{Z}_{d_l}^{+N-M_l}}$ defined by Eq.(\ref{eq:def_POVM_K}), and sent the measurement outcomes $\vec{y}_l$ to all terminal nodes. 

    \item[Step 2] Kobayashi et al.'s multi-source multicast QNC protocol is implemented on $G$
    by choosing the symmetric subspace of  $\mathcal{S}_l$ on $s_l$ as an initial system and 
    $\mathcal{A}_n^{(l)}$ on $t_n$ as a target system for all $l\in [L]$ and $n\in [N]$. 

    \item[Step 3-1] The terminal nodes prepare $\ket{\Upsilon}$ defined by Eq.(\ref{eq def Upsilon}) on $\mathcal{D}^{(l)}\otimes \big( \bigotimes _{n=1}^N \mathcal{C}_n^{(l)}\big)$, $\ket{\vec{y}_l}$ on $\mathcal{G}_n^{(l)}$, $\ket{\vec{0}^{N}}$ on $\mathcal{F}^{(l)}$, and $\ket{0}$ on $\mathcal{E}_n^{(l)}$ for all $l\in [L]$ and $n\in [N]$, where $\vec{0}^{N}$ is a sequence of $N$ zeros. 

    \item[Step 3-2]  
    Terminal node $t_1$ applies $V_0^{(l)}$ on $\mathcal{A}_1^{(l)}\otimes \mathcal{G}_1^{(l)}\otimes\mathcal{D}^{(l)}\otimes\mathcal{F}^{(l)}$ for all $l\in [l]$, and 
    terminal nodes $t_n$ apply $V_n^{(l)}$ on $\mathcal{A}_n^{(l)}\otimes\mathcal{G}_n^{(l)}\otimes\mathcal{C}_n^{(l)}\otimes\mathcal{E}_n^{(l)}$ for all $l\in [l]$ and $n\in [N]$.

    \item[Step 3-3] Terminal nodes remove $\mathcal{A}_n^{(l)}\otimes \mathcal{G}_n^{(l)} \otimes \mathcal{C}_n^{(l)}$ by using $\mathcal{A}_1^{(l)}\otimes \mathcal{G}_1^{(l)} \otimes \mathcal{D}^{(l)}$ with the LOCC subroutine for all $2 \le n \le N$ and $l\in[L]$. Further, terminal node $t_1$ removes $\mathcal{C}_1^{(l)}$ by using $\mathcal{D}^{(l)}$ with the LOCC subroutine for all $l\in [L]$.

\item[Step 3-4]  Terminal node $t_1$ removes $\mathcal{A}_1^{(l)}\otimes \mathcal{G}_1^{(l)} \otimes \mathcal{D}^{(l)}$ by using $\mathcal{F}^{(l)}$ with the LOCC subroutine for all $l\in [L]$.

    \item[Step3-5]  Terminal node $t_1$ removes $\mathcal{F}^{(l)}$ by using $\bigotimes_{n\in [N]} \mathcal{E}_n^{(l)}$ with the LOCC subroutine for all $l\in [L]$.
\end{description}
After Step 3-5, optimal symmetric universal clones of $\ket{\Psi_l}$ are found on $\bigotimes_{n\in [N]}\mathcal{E}_n^{(l)}$.

\subsection{Proof of the main theorem}
In this subsection, we give a proof of the main theorem given in \ref{subsec main theorem}. 
That is, we show that under the assumption given in \ref{subsec main theorem}, the protocol given in \ref{subsec protocol} achieves the multicast of symmetric universal clones of the input state $\ket{\Psi_l}$ from $s_l$ to $t_n$ for all $l\in [L]$ and $n\in [N]$.

We first assume that the input state $\ket{\Psi_l}$ is written as 
$\ket{\Psi_l}=\sum_{a\in \mathbb{Z}_{d_l}}\alpha_a^{(l)}\ket{a}$ for all $l\in [L]$.
Then, for all $l\in [L]$, $\ket{\Psi_l}^{\otimes M_l}\in \mathcal{S}_l$ is written as
\begin{align}\label{eq ket Psi l otimes M l}
\ket{\Psi_l}^{\otimes M_l}=\sum_{\vec{a}\in \mathbb{Z}_{d_l}^{+M_l}}\alpha_{\vec{a}}^{(l)}\ket{\vec{a}}_{sym},    
\end{align}
where $\ket{\vec{a}}_{sym}$ is defined by Eq.(\ref{eq:def ket vec z sym}), and $\alpha_{\vec{a}}^{(l)}$ is defined by 
\begin{align}\label{eq def alpha vec a}
    \alpha_{\vec{a}}^{(l)}:=|\Omega_{\vec{a}}|^{\frac{1}{2}}\prod _{m=1}^M \alpha_{a_m}^{(l)}.
\end{align}
$\Omega_{\vec{a}}$ is defined by Eq.(\ref{eq def Omega}).
If $\mathcal{S}_l$ is measured by $\{K_{\vec{y}_l}\}_{\vec{y}_l\in\mathbb{Z}_{d_l}^{+N-M_l}}$ and derive the measurement outcome $\vec{y_l}$ in the step 2, 
then, $K_{\vec{y}_l} \ket{\Psi_l}^{\otimes M_l}/\|K_{\vec{y}_l} \ket{\Psi_l}^{\otimes M_l}\|$ is derived in probability $\|K_{\vec{y}_l} \ket{\Psi_l}^{\otimes M_l}\|^2$. 
Here, we focus on the unnormalized state $K_{\vec{y}_l} \ket{\Psi_l}^{\otimes M_l}$, which can be written as 
\begin{align}\label{eq state after step 1}
K_{\vec{y}_l} \ket{\Psi_l}^{\otimes M_l}=\sum_{\vec{a}\in \mathbb{Z}_{d_l}^{+M_l}}\widetilde{\alpha}_{\vec{a}}^{(l)}\ket{\vec{a}}_{sym},    
\end{align}
where $\widetilde{\alpha}_{\vec{a}}^{(l)}$ is defined by 
\begin{align}\label{eq def widetilde alpha vec a}
    \widetilde{\alpha}_{\vec{a}}^{(l)}
    :=\sqrt{\frac{d_l[M_l]|\Omega_{\vec{y}_l}|}{d_l[N]|\Omega_{\vec{a}\vec{y_l}}|}}
    \alpha_{\vec{a}}^{(l)}.
\end{align}
In order to treat probability and states togerther, we will continue putting $\|K_{\vec{y}_l} \ket{\Psi_l}^{\otimes M_l}\|$, which is equal to the square root of the probability of obtaining the measurement outcome $\vec{y}_l$, on the states in the following analysis.
 We can easily see that $K_{\vec{y}_l} \ket{\Psi_l}^{\otimes M_l}$ is in the symmetric subspace of 
$\mathcal{S}_l$ for all $l\in [L]$. We should note that the dimension of this subspace is $d_l[M_l]$.

Since we have assumed that there exists an acyclic directed graph $G'$ such that $G'$ is derived by adding directions on $G$, and 
multi-source multicast classical network coding problem on acyclic directed graph $G'$ is solvable with rate $r_l=\lceil log_q  d[M_l] \rceil$, Kobayashi et al's multi-source multicast QNC protocol can be 
implemented on $G$ by choosing the symmetric subspace of $\mathcal{S}_l$ as input systems for all $l\in [L
]$ of in the Step 2.

After Kobayashi et al.'s protocol in the Step 2, the state on $\bigotimes _{l\in[L]}\bigotimes _{n\in[N]}\mathcal{A}_n^{(l)}$ can be written as 
\begin{equation}
 \big(  \sum_{\vec{a}\in \mathbb{Z}_{d_1}^{+M_1}}\widetilde{\alpha}_{\vec{a}}^{(1)}\bigotimes _{n\in[N]} \ket{\vec{a}}_{\mathcal{A}_n^{(1)}} \big)\otimes \cdots \otimes 
  \big(  \sum_{\vec{a}\in \mathbb{Z}_{d_L}^{+M_L}}\widetilde{\alpha}_{\vec{a}}^{(L)}\bigotimes _{n\in[N]} \ket{\vec{a}}_{\mathcal{A}_n^{(L)}} \big).
\end{equation}
After step 2 until the end of the protocol, 
the quantum information processing on $\mathcal{A}_n^{(l)}$, $\mathcal{C}_n^{(l)}$, $\mathcal{E}_n^{(l)}$, 
 $\mathcal{G}_n^{(l)}$, $\mathcal{D}^{(l)}$, $\mathcal{F}^{(l)}$ is independently implemented with respect to $l$. 
Hence, we will omit the phrase ``{\it for all $l\in [L]$}''. 

By means of Eq.(\ref{eq def Upsilon}), 
the state on $\mathcal{D}^{(l)}\otimes \mathcal{F}^{(l)} 
\bigotimes_{n\in [N]} (\mathcal{A}_n^{(l)}\otimes \mathcal{G}_n^{(l)}\otimes\mathcal{C}_n^{(l)}\otimes\mathcal{E}_n^{(l)})$ after step 3-1 can be written as 
\begin{equation}
    \sum_{\vec{a}\in \mathbb{Z}_{d_l}^{+M_l}} \frac{\widetilde{\alpha_{\vec{a}}}^{(l)}}{\sqrt{N!}}
    \sum_{\sigma \in S_N} \ket{\sigma,\vec{0}^N}_{\mathcal{D}^{(l)}\mathcal{F}^{(l)}}\otimes \big( \bigotimes _{n\in[N]} \ket{\vec{a},\vec{y}_l,\sigma(n),0}_{\mathcal{A}_n^{(l)}\mathcal{G}_n^{(l)}\mathcal{C}_n^{(l)}\mathcal{E}_n^{(l)}} \big).
\end{equation}
Then, by means of Eqs.(\ref{eq def V 0 l}) and (\ref{eq def V n l}), the state after step 3-2 can be written as 
\begin{equation}
    \sum_{\vec{a}\in \mathbb{Z}_{d_l}^{+M_l}} \frac{\widetilde{\alpha_{\vec{a}}}^{(l)}}{\sqrt{N!}}
    \sum_{\sigma \in S_N} \ket{\sigma,\sigma(\vec{a}\vec{y}_l)}_{\mathcal{D}^{(l)}\mathcal{F}^{(l)}}\otimes \big( \bigotimes _{n\in[N]} \ket{\vec{a},\vec{y}_l,\sigma(n),(\sigma(\vec{a}\vec{y}_l))_n}_{\mathcal{A}_n^{(l)}\mathcal{G}_n^{(l)}\mathcal{C}_n^{(l)}\mathcal{E}_n^{(l)}} \big),
\end{equation}
where $\sigma(\vec{a}\vec{y}_l)$ is defined by Eq.(\ref{eq def sigma vec z}).
Here,  $(\sigma(\vec{a}\vec{y}_l))_n$ is the n-th entry of $\sigma(\vec{a}\vec{y}_l)$, which is equal to the $\sigma(n)$-th entry of $\vec{a}\vec{y}_l$.

Since $(\vec{a},\vec{y}_l,\sigma(n))$ is uniquely determined by $(\vec{a},\vec{y}_l,\sigma)$,
the terminal nodes can remove $\mathcal{A}_n^{(l)}\otimes \mathcal{G}_n^{(l)} \otimes \mathcal{C}_n^{(l)}$ by using $\mathcal{A}_1^{(l)}\otimes \mathcal{G}_1^{(l)} \otimes \mathcal{D}^{(l)}$ with the LOCC subroutine for all $2 \le n \le N$ and $l\in[L]$. 
Further, since $\sigma(1)$ is uniquely determined by $\sigma\in S_N$,
terminal node $t_1$ can removes $\mathcal{C}_1^{(l)}$ by using $\mathcal{D}^{(l)}$ with the LOCC subroutine for all $l\in [L]$.
Then, the state after step 3-3 can be written as 
\begin{equation}\label{eq state step 3 3}
    \sum_{\vec{a}\in \mathbb{Z}_{d_l}^{+M_l}} \frac{\widetilde{\alpha_{\vec{a}}}^{(l)}}{\sqrt{N!}}
    \sum_{\sigma \in S_N} \ket{\vec{a},\vec{y}_l,\sigma,\sigma(\vec{a}\vec{y}_l)}_{\mathcal{A}_1^{(l)}\mathcal{G}_1^{(l)}\mathcal{D}^{(l)}\mathcal{F}^{(l)}}\otimes \big( \bigotimes _{n\in[N]} \ket{(\sigma(\vec{a}\vec{y}_l))_n}_{\mathcal{E}_n^{(l)}} \big).
\end{equation}

To show that we can implement step 3-4, we need to use the following relation:
\begin{equation}\label{eq sum sigma in S N ket g}
    \sum_{\sigma\in S_N} \ket{g(\vec{c},\sigma,\sigma \vec{c})}
    =\sum_{\vec{b}\in \Omega_{\vec{c}}} \sum_{\sigma \in \overline{\Omega}_{\vec{c},\vec{b}}}  \ket{g(\vec{c},\sigma,\vec{b})},
\end{equation}
where $\overline{\Omega}_{\vec{c},\vec{b}}$ is a subset of $S_N$ defined by 
\begin{equation}
    \overline{\Omega}_{\vec{c},\vec{b}}:= \{\sigma \in S_N\ | \ \sigma \vec{c} = \vec{b} \}.
\end{equation}
We can show that if $\overline{\Omega}_{\vec{c},\vec{b}}\neq \emptyset$, then, $\overline{\Omega}_{\vec{c},\vec{b}}$ is
a right coset of $H_{\vec{b}}$; see appendix A.
We can easily see that Eq.(\ref{eq sum sigma in S N ket g}) holds for any vector $\ket{g(\vec{c},\sigma,\vec{b})}$.

By using Eq.(\ref{eq sum sigma in S N ket g}) with $\vec{c}=\vec{a}\vec{y}_l$, the state given by Eq.(\ref{eq state step 3 3}) 
can be rewritten as 
\begin{equation}\label{eq sum vec in a mathbb Z d l M l}
    \sum_{\vec{a}\in \mathbb{Z}_{d_l}^{+M_l}} 
    \sum_{\vec{b} \in \Omega_{\vec{a}\vec{y}_l}} 
    \widetilde{\alpha_{\vec{a}}}^{(l)} \sqrt{\frac{|\overline{\Omega}_{\vec{a}\vec{y}_l,\vec{b}}|}{N!}}\ket{\psi_{\vec{a}\vec{y}_l}(\vec{b})}_{\mathcal{A}_1^{(l)}\mathcal{G}_1^{(l)}\mathcal{D}^{(l)}}\ket{
    \vec{b}}_{\mathcal{F}^{(l)}}\otimes \big( \bigotimes _{n\in[N]} \ket{b_n}_{\mathcal{E}_n^{(l)}} \big),
\end{equation}
where $b_n$ is $n$-th entry of $\vec{b}$.
Here, for $\vec{a}\in \mathbb{Z}_{d_l}^{+M_l},\vec{y}_l\in \mathbb{Z}_{d_l}^{+N-M_l}$ and $\vec{b}\in \Omega_{\vec{a}\vec{y_l}}$, $\ket{\psi_{\vec{a}\vec{y}_l}(\vec{b})}$ is defined by 
\begin{equation}\label{eq def psi vec a vec y vec b}
    \ket{\psi_{\vec{a}\vec{y}_l}(\vec{b})}:=|\overline{\Omega}_{\vec{a}\vec{y}_l,\vec{b}}|^{-\frac{1}{2}}
    \ket{\vec{a},\vec{y}_l}_{\mathcal{A}^{(l)},\mathcal{G}_1^{(l)}}
    \sum_{\sigma \in \overline{\Omega}_{\vec{a}\vec{y}_l,\vec{b}}} \ket{\sigma}_{\mathcal{D}^{(l)}}.
\end{equation}
Eq.(\ref{eq def psi vec a vec y vec b}) leads that, for a given $\vec{y}_l\in \mathbb{Z}_{d_l}^{+N-M_l}$, $\{ \ket{\psi_{\vec{a}\vec{y}_l}(\vec{b})} \}_{\vec{a}\in \mathbb{Z}_{d_l}^{+M_l},\vec{b}\in \Omega_{\vec{a}\vec{y_l}}}$ is an orthonormal set.  
Further, for all $\vec{y}_l\in \mathbb{Z}_{d_l}^{+N-M_l}$ and  $\vec{a}, \vec{a'}\in \mathbb{Z}_{d_l}^{+M_l}$, $\vec{a} \neq \vec{a'}$ leads $\Omega_{\vec{a}\vec{y_l}}\cap \Omega_{\vec{a'}\vec{y_l}}=\emptyset$. 
Suppose $\widetilde{\Omega}_{\vec{y_l}}$ is defined by 
\begin{equation}
    \widetilde{\Omega}_{\vec{y_l}}:= \bigcup_{\vec{a}\in \mathbb{Z}_{d_l}^{+M_l}} \Omega_{\vec{a}\vec{y_l}}.
\end{equation}
Then, for a given $\vec{b}\in \widetilde{\Omega}_{\vec{y_l}}$, there exists a unique $\vec{a}\in \mathbb{Z}_{d_l}^{+M_l}$ satifying  $\vec{b}\in \Omega_{\vec{a}\vec{y_l}}$.
Hence, we can define a map $\theta: \widetilde{\Omega}_{\vec{y_l}}\rightarrow \mathbb{Z}_{d_l}^{+M_l}$ such that 
$\vec{b}\in \Omega_{\theta (\vec{b})\vec{y_l}}$. By using $\theta$, we can rewrite the state given by Eq.(\ref{eq sum vec in a mathbb Z d l M l}) as 
\begin{equation}\label{eq sum vec b in Omega}
    \sum_{\vec{b} \in \widetilde{\Omega}_{\vec{y}_l}} 
    \widetilde{\alpha_{\theta (\vec{b})}}^{(l)} \sqrt{\frac{|\overline{\Omega}_{\theta (\vec{b})\vec{y}_l,\vec{b}}|}{N!}}\ket{\psi_{\theta (\vec{b})\vec{y}_l}(\vec{b})}_{\mathcal{A}_1^{(l)}\mathcal{G}_1^{(l)}\mathcal{D}^{(l)}}\ket{
    \vec{b}}_{\mathcal{F}^{(l)}}\otimes \big( \bigotimes _{n\in[N]} \ket{b_n}_{\mathcal{E}_n^{(l)}} \big).
\end{equation}
Now, since $\{ \ket{\psi_{\theta (\vec{b})\vec{y}_l}(\vec{b})}\}_{\vec{b} \in \widetilde{\Omega}_{\vec{y}_l}}$ is an orthonormal set, it is clear that 
terminal node $t_1$ can remove $\mathcal{A}_1^{(l)}\otimes \mathcal{G}_1^{(l)} \otimes \mathcal{D}^{(l)}$ by using $\mathcal{F}^{(l)}$ with the LOCC subroutine for all $l\in [L]$ at the step 3-4.

Eq.(\ref{eq sum vec b in Omega}) leads that the state on $\mathcal{F}^{(l)}\otimes \big( \bigotimes _{n\in[N]} \ket{b_n}_{\mathcal{E}_n^{(l)}} \big)$ after the step 3-4 can be written as 
\begin{equation}\label{eq sum vec b in Omega 2}
    \sum_{\vec{b} \in \widetilde{\Omega}_{\vec{y}_l}} 
    \widetilde{\alpha_{\theta (\vec{b})}}^{(l)} \sqrt{\frac{|\overline{\Omega}_{\theta (\vec{b})\vec{y}_l,\vec{b}}|}{N!}}\ket{
    \vec{b}}_{\mathcal{F}^{(l)}}\otimes \big( \bigotimes _{n\in[N]} \ket{b_n}_{\mathcal{E}_n^{(l)}} \big).
\end{equation}
Now, we can define $d_l$-dimensional sytem $\mathcal{F}_n^{(l)}$ for $n\in[N]$ satisfying 
$\mathcal{F}^{(l)}=\bigotimes_{n\in [N]}$ and 
\begin{equation}
    \ket{    \vec{b}}_{\mathcal{F}^{(l)}}=\ket{
    b_1}_{\mathcal{F}_1^{(l)}}\otimes \cdots \otimes \ket{
    b_N}_{\mathcal{F}_N^{(l)}},
\end{equation}
where $\vec{b}=(b_1,\cdots , b_N)$.
Then, the state given by Eq.(\ref{eq sum vec b in Omega 2}) can be rewritten as 
\begin{equation}\label{eq sum vec b in Omega 3}
    \sum_{\vec{b} \in \widetilde{\Omega}_{\vec{y}_l}} 
    \widetilde{\alpha_{\theta (\vec{b})}}^{(l)} \sqrt{\frac{|\overline{\Omega}_{\theta (\vec{b})\vec{y}_l,\vec{b}}|}{N!}}\big( \bigotimes _{n\in[N]} \ket{b_n}_{\mathcal{F}_n^{(l)}} \big)\otimes \big( \bigotimes _{n\in[N]} \ket{b_n}_{\mathcal{E}_n^{(l)}} \big).
\end{equation}
Thus, it is clear that 
$\mathcal{F}_n^{(l)}$ can be removed $\mathcal{E}_n^{(l)}$
 with the LOCC subroutine for all $n\in [N]$ at step 3-5.
Hence, the state after step 3-5 is 
 \begin{align}\label{eq sum vec b in Omega 4}
&    \sum_{\vec{b} \in \widetilde{\Omega}_{\vec{y}_l}} 
    \widetilde{\alpha_{\theta (\vec{b})}}^{(l)} \sqrt{\frac{|\overline{\Omega}_{\theta (\vec{b})\vec{y}_l,\vec{b}}|}{N!}} \big( \bigotimes _{n\in[N]} \ket{b_n}_{\mathcal{E}_n^{(l)}} \big)
    \nonumber \\
=    & 
    \sum_{\vec{a}\in \mathbb{Z}_{d_l}^{+M_l}} 
    \sum_{\vec{b} \in \Omega_{\vec{a}\vec{y}_l}} 
    \widetilde{\alpha_{\vec{a}}}^{(l)} \sqrt{\frac{|\overline{\Omega}_{\vec{a}\vec{y}_l,\vec{b}}|}{N!}} \big( \bigotimes _{n\in[N]} \ket{b_n}_{\mathcal{E}_n^{(l)}} \big)\nonumber \\
    =    & 
    \sum_{\vec{a}\in \mathbb{Z}_{d_l}^{+M_l}} 
         \widetilde{\alpha_{\vec{a}}}^{(l)}  
    \frac{1}{\sqrt{\Omega_{\vec{a}\vec{y}_l}}}\big( \sum_{\vec{b} \in \Omega_{\vec{a}\vec{y}_l}} \bigotimes _{n\in[N]} \ket{b_n}_{\mathcal{E}_n^{(l)}} \big)\nonumber \\
    =    & 
    \sum_{\vec{a}\in \mathbb{Z}_{d_l}^{+M_l}} 
\sqrt{\frac{d_l[M_l]\cdot |\Omega_{\vec{y}_l}|\cdot |\Omega_{\vec{a}}|}{d_l[N]\cdot |\Omega_{\vec{a}\vec{y_l}}|}}
    \prod _{m=1}^M \alpha_{a_m}^{(l)}
\ket{\vec{a}\vec{y}}_{sym} \nonumber \\
    =    & 
V_{\vec{y}_l}K_{\vec{y}_l}\ket{\Psi_l}^{\otimes M_l}.
\end{align}
Here, the relation $|\overline{\Omega}_{\vec{x},\vec{y}}|=\frac{N!}{|\Omega_{\vec{x}}|}$
given in Appendix A is used in the second equality, Eqs.(\ref{eq def alpha vec a}) and (\ref{eq def widetilde alpha vec a}) are used in the third equality, and Eqs. (\ref{eq:def_isometry_V}),(\ref{eq:def_POVM_K}) and (\ref{eq ket Psi l otimes M l}) are used in the forth equality. 

Since we have implemented a measurement on $K_{\vec{y}}$, and continue calculating the unnormalized state. 
Eqs.(\ref{eq:uqcm_K_V}) and (\ref{eq sum vec b in Omega 4}) shows that the state on $\bigotimes_{n\in [N]}\mathcal{E}_n^{(l)}$ after the protocol is $\sum_{\vec{y}\in \mathbb{Z}_{d_l}^{+N-M_l}}P(V_{\vec{y}_l}K_{\vec{y}_l}\ket{\Psi_l}^{\otimes M_l})=\Lambda_{UQCM}(P(\ket{\Psi_l}^{\otimes M_l}))$.
Thus, the multi-source multicast QNC problem on $G$ for symmetric universal cloning with $L$ source nodes, $N$ terminal nodes, and $M_l$ copies of $d_l$-dimensional system for all $l\in [L]$ is solvable.
We have proven the main theorem.

\section{Summary}\label{sec summary}
In this paper, we studied the problem of perfectly multicasting optimal symmetric universal quantum clones over quantum networks with free classical communication.
 We constructed a protocol that multicasts symmetric universal clones of input states from multiple source nodes to multiple target nodes by extending the quantum network coding protocol proposed by Kobayashi et al. \cite{Kobayashi2010}.

 We first established a sufficient condition for perfect multicast in the general multi-source setting.  
 Specifically, we showed that for source nodes $s_1, \ldots, s_L$, where $M_l$ copies of a $d_l$-dimensional input state $\ket{\Psi_l}$ are available at each source node $s_l$, perfect multicast of the corresponding symmetric universal clones to all target nodes is achievable under the assumption that a small amount of entanglement is shared among the target nodes. 
This is possible if there exists a classical linear multi-source multicast network code over $\mathbb{F}_q$ on a directed graph $G'$ that achieves multicast rates $\lceil \log_q d_l [M_l] \rceil$ for all $l$.
  Here, the directed graph $G'$ is obtained by assigning directions to the edges of the underlying undirected quantum network $G$, where each edge corresponds to a noiseless $q$-dimensional quantum channel, and $d[M]:=\frac{(d+M-1)!}{M!(d-1)!}$ denotes the dimension of the symmetric subspace of $M$ copies of a $d$-dimensional system.

Taking into account the max-flow min-cut theorem for classical multicast network coding, the above result can be rephrased as follows for the single-source setting:
When a single copy of a $q^r$-dimensional input state is available at a source node $s$, where $q$ is a sufficiently large prime power, perfect multicast of the symmetric universal clone is achievable using a small amount of entanglement shared among the target nodes. This holds provided that the undirected quantum network $G$ admits an acyclic directed graph $G'$ such that all minimum cuts between the source node $s$ and each target node are at least $r$. In other words, the achievable multicast rate of symmetric universal clones is characterized by the minimum cut of the quantum network.

\section*{Appendix A}

In this appendix, we introduce properties of the symmetric group $S_N$
acting on $\mathbb{Z}_{d}^N$ and the symmetric subspace of $\mathbb{C}_{d}^{\otimes N}$
which are necessary in the proofs of the main text.

Suppose a stabilizer subgroup $H_{\vec{z}}\subset S_N$ for $\vec{z}\in\mathbb{Z}_{d}^N$
is defined as $H_{\vec{z}}:=\left\{ \sigma\in S_N|\sigma\vec{z}=\vec{z}\right\} $.
Then, for all $\sigma,$$\sigma'\in S_N$, 
\[
\sigma'\in\sigma H_{\vec{z}}\Longleftrightarrow\sigma^{-1}\sigma'\in H_{\vec{z}}\Longleftrightarrow\sigma\vec{z}=\sigma'\vec{z}.
\]

Hence, the index $\left[S_N:H_{\vec{z}}\right]$, the number of
the left cosets of $H_{\vec{z}}$ in $S_N$, is equal to $|\Omega_{\vec{z}}|$,
where $\Omega_{\vec{z}}:=\left\{ \sigma\vec{z}\ |\ \sigma\in S_N\right\} $. This
fact and Lagrange's theorem of finite groups lead
\[
N!=\left|S_N\right|=\left[S_N:H_{\vec{z}}\right]\cdot\left|H_{\vec{z}}\right|=\left|\Omega_{\vec{z}}\right|\cdot\left|H_{\vec{z}}\right|.
\]
The above equation implies $\sum_{\sigma\in S_N}\ket{\sigma\vec{x}}=\left|H_{\vec{x}}\right|\cdot\sum_{\vec{x}'\in\Omega_{\vec{x}}}\ket{\vec{x}'}$.
Thus, $\left|\vec{x}\right\rangle _{sym}$ defined by \eqref{eq:def ket vec z sym}
is a unit vector. For any $\vec{x},$ $\vec{x}'\in\mathbb{Z}_{d}^N$,
$\left|\vec{x}\right\rangle _{sym}=\left|\vec{x}'\right\rangle _{sym}$
if and only if $\vec{x}'$ can be derived from $\vec{x}$ by a permutation.
In other words, the following relation holds:
\[
\left|\vec{x}\right\rangle _{sym}=\left|\vec{x}'\right\rangle _{sym}\Longleftrightarrow\exists\sigma\in S_N,\ s.t.\ \vec{x}'=\sigma\vec{x}\Longleftrightarrow\Omega_{\vec{x}}=\Omega_{\vec{x}'}.
\]

For given $\vec{x}, \vec{y}\in \mathbb{Z}_{d}^N$, we define $\overline{\Omega}_{\vec{x},\vec{y}}=\{ \sigma \in S_N \ | \ \sigma \vec{x}=\vec{y}\}$. Suppose $\overline{\Omega}_{\vec{x},\vec{y}}\neq \emptyset$.
Then, we derive the following relation:
\begin{equation}\label{eq appendix sigma}
    \sigma \in \overline{\Omega}_{\vec{x},\vec{y}}, \sigma'\in H_{\vec{y}}
    \Rightarrow \sigma' \sigma \vec{x}=\vec{y}\Rightarrow \sigma' \sigma \in  \overline{\Omega}_{\vec{x},\vec{y}}.
\end{equation}
Further, we derive the following relation:
\begin{equation}\label{eq appendix sigma 1}
    \sigma_1, \sigma_2\in \overline{\Omega}_{\vec{x},\vec{y}} \Rightarrow \vec{x}=\sigma_1^{-1}\vec{y}=\sigma_2^{-1}\vec{y} \Rightarrow \sigma_1 \sigma_2^{-1}\in H_{\vec{y}}.
\end{equation}

Suppose $\sigma\in \overline{\Omega}_{\vec{x},\vec{y}}$. Then, Eqs.(\ref{eq appendix sigma}) and (\ref{eq appendix sigma 1}) guarantee \begin{equation}
    \sigma'\in \overline{\Omega}_{\vec{x},\vec{y}}\Leftrightarrow \sigma'\sigma^{-1}\in H_{\vec{y}} \Leftrightarrow \sigma'\in H_{\vec{y}}\sigma.
\end{equation}
Therefore, in the case $\overline{\Omega}_{\vec{x},\vec{y}}\neq \emptyset$, we have the relation $\overline{\Omega}_{\vec{x},\vec{y}}=H_{\vec{y}}\sigma$ for an arbitrary $\sigma \in \overline{\Omega}_{\vec{x},\vec{y}}$; that is $\overline{\Omega}_{\vec{x},\vec{y}}$ is a right coset of $H_{\vec{y}}$.
Thus, we have $|\overline{\Omega}_{\vec{x},\vec{y}}|=|H_{\vec{y}}|=\frac{N!}{|\Omega_{\vec{y}}|}$.
Further, $\overline{\Omega}_{\vec{x},\vec{y}}^{-1}:=\{\sigma^{-1}\ |\ \sigma \in  \overline{\Omega}_{\vec{x},\vec{y}} \}$ satisfies $\overline{\Omega}_{\vec{x},\vec{y}}^{-1}=\overline{\Omega}_{\vec{y},\vec{x}}$.
This relation leads $|\overline{\Omega}_{\vec{x},\vec{y}}|=|\overline{\Omega}_{\vec{y},\vec{x}}^{
-1}|=|\overline{\Omega}_{\vec{y},\vec{x}}|=|H_{\vec{x}}|=\frac{N!}{|\Omega_{\vec{x}}|}$.

\paragraph{Acknowledgment}
G.K. acknowledges the support from JST CREST (Grant
Number JPMJCR2113, Japan) and JSPS KAKENHI (Grant Number JP23K25793).
M.O. acknowledges the support from JSPS KAKENHI (Grant Number JP24H00952).
This work was the Project for Developing Innovation Systems of MEXT, Japan, and 
JSPS KAKENHI (Grants No. 26330006 and No. 16H01050).

%
%
\bibliographystyle{unsrt}
\bibliography{secure_network_coding}
\end{document}